# Complementary ab initio and X-ray nano-diffraction study of Ta$_2$O$_5$

R. Hollerweger[1], D. Holec[2], J. Paulitsch[3], M. Bartosik[1], R. Daniel[2], R. Rachbauer[4], P. Polcik[5], J. Keckes[6], C. Krywka[7], P. H. Mayrhofer[1,3]

[1] Christian Doppler Laboratory for Application Oriented Coating Development at the Institute of Materials Science and Technology, Vienna University of Technology, A-1040 Vienna, Austria

[2] Department of Physical Metallurgy and Materials Testing, Montanuniversität Leoben, A-8700 Leoben, Austria

[3] Institute of Materials Science and Technology, Vienna University of Technology, A-1040 Vienna, Austria

[4] Oerlikon Balzers Coating AG, LI-9469 Balzers, Principality of Liechtenstein

[5] Plansee Composite Materials GmbH, D-86983 Lechbruck am See, Germany

[6] Department Materials Physics, Montanuniversität Leoben and Materials Center Leoben, A-8700 Leoben, Austria

[7] Ruprecht Haensel Laboratory, University of Kiel, Leibnizstrasse 19, D-24098 Kiel and Helmholtz-Zentrum Geesthacht, Institute for Materials Research, D-21502 Geesthacht, Germany

robert.hollerweger@tuwien.ac.at

**Abstract**

Numerous different crystal structures of Ta$_2$O$_5$ are reported in literature. Although experimentally and computationally obtained lattice parameters and mechanical properties are in excellent agreement there is a pronounced deviation when it comes to electronic structures of Ta$_2$O$_5$. Based on *ab initio* studies and nano-beam X-ray diffraction of sputtered Ta$_2$O$_5$ thin films, we introduce an orthorhombic basic structure with a = 6.425 Å, b = 3.769 Å, and c = 7.706 Å, which is stabilized by flipping of an oxygen atom in neighboring c-planes. The calculated energy of formation is with −3.209 eV/atom almost as low as −3.259 eV/atom for the well-known Stephenson superstructure. We propose the new structure based on the fact that it allows for a good description of orthorhombic $Ta_2O_5$ even with a small and simple unit cell, which is especially advantageous for computational studies.

1. **Introduction**

Tantalum pentoxide ($Ta_2O_5$) is an important semiconductor material exhibiting a high dielectric constant and refractive index [1,2] and therefore it is commonly used for capacitors and optical coatings. Moreover, it is a highly biocompatible [3] and corrosion resistant [2,4] material and recent investigations also indicate piezoelectric properties [5,6].

The crystal structure of $Ta_2O_5$ has been carefully studied for more than 50 years. As one of the first, Lehovec [7] has attributed an orthorhombic structure to tantalum pentoxide, which is nowadays known as β-$Ta_2O_5$. Furthermore, he observed some additional peaks within his X-Ray Diffraction (XRD) patterns which probably originate from superstructures like described by Grey et al. [8], Audier et al. [9] or Stephenson et al. [10]. Investigations on the thermal stability of these coatings by Lagergren et al. [11] have shown a phase transformation of the orthorhombic β-$Ta_2O_5$ to a tetragonal high temperature phase (α-$Ta_2O_5$) at around 1320 °C. Moreover, a hexagonal metastable low temperature phase (δ-$Ta_2O_5$) is reported [12], which irreversibly transforms into β-$Ta_2O_5$ at elevated temperatures. A very characteristic property of $Ta_2O_5$ low temperature structures is the O position, which is exactly between two tantalum atoms in c-direction for most of the structures. Contrary, the O distribution in a- and b-direction may vary significantly. This may be responsible for the existence of all the different reported orthorhombic, hexagonal, monoclinic or triclinic structures as well as for the superstructures. For more details we refer to a summary given by Hummel et al.[13].

Most of the reported crystal structures are closely related to each other. A simple monoclinic cell with angles of γ = 120 deg. is equal to a hexagonal cell, which itself contains a base centered orthorhombic lattice with an a/b ratio of $1/\sqrt{3}$. Moreover, $Ta_2O_5$ can also form the tetragonal structure, as the b and c lattice constants of the orthorhombic structure are not too different (0.366 nm vs. 0.388 nm). This clearly indicates that small variations of the oxygen distribution within the a-b plane can significantly influence the resulting structure of $Ta_2O_5$.

Based on cross sectional X-ray nano-diffraction of reactively sputtered $Ta_2O_5$ thin films, and Density Functional Theory (DFT) studies we introduce an orthorhombic $Ta_2O_5$ structure, which allows for an easier description as compared to the superstructures. Our results contribute to the understanding of the formation of various types of low temperature β- and δ-$Ta_2O_5$ and the superstructures.

## 2. Experimental:

Tantalum oxide films were synthesized in a laboratory scale Leybold Heraeus magnetically unbalanced magnetron sputtering device. Therefore, a metallic tantalum target (99.95% purity) with a diameter of 75 mm was mounted to the copper cathode and DC sputtered in an $Ar/O_2$ glow discharge with a flow ratio of 50 % at a total pressure of 0.4 Pa. The depositions were performed on (100) oriented silicon substrates (20 x 7 x 0.35 mm³). Detailed information on the deposition conditions and the experimental setup can be found in Ref. [14].

Vacuum annealing treatments were performed at a base pressure of at least $5 \times 10^{-4}$ Pa, with a heating rate of 20 K/min, a holding time of 1 h at up to 1000 °C, and a cooling rate of 50 K/min. Post oxidation of our films was carried out in ambient air for 1 h at 1000 °C in a Nabertherm furnace.

The elemental composition was investigated by Elastic Recoil Detection Analysis (ERDA) using 35 MeV $Cl^{7+}$ incident ions. For obtaining a quantitative analysis, the resulting elemental spectrum was evaluated after Barrada et al. [15].

An evaluation of the mechanical properties was carried out by nanoindentation testing using a UMIS indentation system by Fischer Cripps Laboratories equipped with a Berkovich indenter tip. The resulting load displacement curves were evaluated after Oliver and Pharr [16] and the used normal loads were ranging between 2 and 10 mN to keep the penetration depths below 10% of the film thickness to avoid substrate interference.

Structural investigations in laboratory conditions were performed in a Bruker D8 X-Ray diffractometer with a CuKα radiation (λ=1.54 Å) and a Sol-X detector in Bragg Brentano geometry. To determine the crystal structure across the layer thickness, cross sectional X-ray nano-diffraction measurements were performed at the nano-focus end-station of P03 Micro and Nanofocus X-ray Scattering (MINAXS) beamline at Petra III synchrotron (Hamburg, Germany). The diameter of the used X-Ray beam was below 500 nm and the wavelength λ was 0.808 Å. Further experimental details on this advanced technique can be found in [17–21]. Surfaces of the coatings and their fracture cross-sectional micrographs were investigated using a Zeiss EVO50 and a FEI XL30 Scanning Electron Microscope (SEM) operating at an acceleration voltage of 15 keV. To avoid surface charging, a thin Au layer was deposited onto the $Ta_2O_5$ cross-sections prior to these investigations.

The structural parameters, energies of formation ($E_f$) and density of states (DOS) of the $Ta_2O_5$ phases were calculated using the Vienna ab initio simulation package (VASP) [22,23] using projector augmented wave pseudopotentials [24] and the generalized gradient approximation (GGA) [25]. The plane wave cut-off energy of 800 eV and more than 2000 k-points·atom ensure a total energy accuracy of 1 meV/atom. Volume, cell-shape and all atom positions were fully optimized during the calculations (for unstable structures the cell shape was fixed).

For visualizing the different crystal structures the program visualization for electronic and structural analysis (VESTA) [26] was used .

## 3. Results and Discussion

### 3.1 Film deposition and characterization

Figure 1 shows an SEM image of the surface (a1) and the fracture cross-section (b1) of reactively sputter deposited $Ta_2O_5$ (deposition time of 5 min). The surface of this thin film is smooth, and the cross-section shows a dense featureless morphology. If the deposition time is increased to 25 min, the microstructure significantly changes (a2) and island-like structures are visible on the surface. The corresponding cross-section (b2) exhibits a smooth, well defined fracture pattern at the substrate-film interface, which becomes significantly disturbed with progressing film thickness. After a deposition time of 120 min, an increased roughening of the film surface is evident (a3) and the cross-sectional micrograph (b3) clearly reveals two different areas. The substrate-film interface near region exhibits a featureless structure, which is overgrown by a fibrous-like structure at increased film thicknesses. The corresponding XRD patterns indicate the presence of an X-ray amorphous structure for the ~0.5 µm thin film (deposition time of 5 min), distinct crystalline peaks for the ~3 µm thin film (deposition time of 25 min), and a predominant crystalline structure for the ~15 µm thick film (deposition time of 120 min), Figs. 1c1-3. These crystalline XRD peaks are indexed according to Lehovec et al. [7,27] and indicate a highly (110) – (200) textured orthorhombic $Ta_2O_5$ phase (note that the intensity axis for the XRD pattern is in logarithmic scale). Moreover, strong shifts of the diffraction peaks of more than 0.5 deg. to lower diffracting angles as compared to the powder diffraction file [7,9] suggest a highly strained lattice in a and b direction. Nevertheless, the XRD spectrum would also be consistent with the hexagonal structure published by Khitrova et al. [28,29] or Fukumoto et al. [30].

Hence, detailed studies of the structural and textural evolution with the film thickness were carried out in 100 nm steps from the substrate-film interface to the film surface by cross-sectional X-ray nano-diffraction using a ~500 nm in diameter X-ray beam. Figure 2 shows a representative 2D diffraction pattern constructed by summing all individual 2D patterns collected across the coating thickness. The sequence of Debye-Scherrer rings can be unambiguously indexed using the orthorhombic $Ta_2O_5$ structure (in good agreement with Figs. 1 c1-c3) whereas no characteristic signs of the hexagonal structure can be detected. The strong variation of the intensity along the diffraction rings suggests a pronounced 110-200 fibre texture for an in-plane isotropic film. The highlighted cake segment of the Debye Scherrer rings at around 0 deg., with only 110 and 200 diffraction rings, corresponds to the diffraction on crystallographic planes oriented approximately parallel to the substrate surface. $\theta$-$2\theta$ XRD laboratory scans (Fig. 1c3) represent diffraction on the equally oriented planes and the results are obviously consistent with those from Fig. 2. The highlighted segment at ~90 deg. representing diffraction on planes oriented perpendicular to the film-substrate interface reveals the presence of all major diffraction planes of the orthorhombic $Ta_2O_5$ structure. Consequently, the XRD results (conventional XRD and nano diffraction analysis) suggest that the films grew with the a-b plane perpendicular to the substrate surface.

Figures 3a, b, and c show the development of the nano-diffraction patterns as a function of the film thickness using the cake-segments at 0, 45 and 90 deg., respectively (the size is defined by an azimuthal angle of 10 deg.). From 0 to about 3 µm only weak diffraction peaks are detectable for the 110 and 200 orientation of the 0 deg. segment and the 001 orientation of the 90 deg. segment. This would disagree with XRD (see the pattern in Fig. 1c1) and with transmission electron microscopy (TEM) investigations (not shown here) which actually indicate an amorphous structure. One explanation is that the laboratory XRD signal originates predominantly from the film surface, whereas nano-diffraction is a local technique. Another possibility is the fact that the Lorentzian profile of the X-ray nano-beam as well as the not-ideal alignment of the beam parallel to the interface caused the intensity increase also for the interface regions. With increasing film thickness from ~3 to ~15 µm (the film surface) the crystalline diffraction peaks become more pronounced. The peak positions for the different orientations well document that the film exhibits an orthorhombic structure across the film thickness, with a pronounced 110 and 200 growth orientations. These patterns can be indexed by the peak positions proposed by Lehovec. Nevertheless, the 110 and 200 peaks collected at 0 deg. and the 111 and 201 peaks of the 45 deg. pattern are highly shifted to lower diffraction angles as indicated by Lehovec. Possible reasons for these off-positions are discussed in the calculation part (section 3.2).

Structural investigations by conventional XRD of our 15 µm thick $Ta_2O_5$ films (deposition time of 120 min) after vacuum as well as ambient air annealing for 1 h at temperatures up to 1000 °C, respectively both indicate the same behavior (Figs. 4a and 4b). With increasing annealing temperature the diffraction peaks approach their reported positions and for $T_a$ = 800 °C additional XRD peaks suggest for superstructure development. The XRD patterns of the films after annealing at 1000 °C exactly match the basic structure reported by Lehovec [7] as well as the 25L superstructure published by Audier et al. [9]. We speculate that the additional (not indexed) peaks at diffracting $2\theta$ angles ≥ 45 deg are also the result of the superstructure, but the corresponding peak positions at these high diffraction angles are not published by Audier et al. Annealing at 1000 °C in vacuum and ambient air causes a pronounced change of the morphology of our coating, compare Fig. 5a and b with Fig. 1b3. Especially the amorphous substrate near region, see Figs. 1, developed a pronounced crystalline-like fracture pattern, in good agreement with results obtained by Wu et al. [31].

The chemical compositions of our ~0.5 µm and ~15 µm (deposition time of 5 and 120 min) clearly indicate a substoichiometric $TaO_x$ with O/Ta ratios of x = 2.39 and 2.33 in the as deposited state, respectively.

The ~0.5 µm thin amorphous coating (Fig. 1a1-1c1) exhibits a hardness (H) of 8.2 ± 0.4 GPa and an indentation modulus (E) of 153 ± 17 GPa, whereas the thicker coating with ~10 µm thick crystalline region at the top has H and E values of 14.3 ± 0.4 and 188 ± 4 GPa. Upon annealing this coating for 1 h at 1000 °C in vacuum the O/Ta ratio decreases from 2.33 to 2.23. The corresponding annealing treatment in ambient air leads to an increase of the O/Ta to 2.57. Irrespective of their slightly different chemical composition upon annealing in vacuum or ambient atmosphere, no difference in the structure can be detected by XRD (Figs. 5a and b). This indicates that the $Ta_2O_5$ superstructures are tolerant for non-stoichiometry. Moreover, also the hardness with 9.3 ± 0.5 and 9.4 ± 0.6 GPa as well as the indentation modulus with 139 ± 6 and 139 ± 3 GPa after annealing at 1000 °C in vacuum and ambient air, respectively, are almost identical. The reduction in hardness and indentation modulus upon annealing can be attributed to recovery and recrystallization effects, resulting in a decreased defect density and stress state [32].

## 3.2 Calculations

Based on these experimental results and in combination to published data we have developed a new $Ta_2O_5$ structure exhibiting an optimized energy of formation ($E_f$) and density of states (DOS). As our as deposited films did not show any superstructure peaks, we have selected several published orthorhombic structures of $Ta_2O_5$, like the subtraction type model based on $U_3O_8$ [33] and the models described by Lehovec [7,33] or Stephenson [10] for comparison. The $U_3O_8$ model is a possible basic structure having one formula unit whereas the structure of Lehovec has two, and the Stephenson model is a superstructure with 11 formula units of $Ta_2O_5$. Our calculations show that the $U_3O_8$ model is not stable in the orthorhombic crystal structure, as the cell shape relaxation yields a monoclinic structure, see Table 1. A full optimization (towards lower energy of formation) of the crystal structure of Lehovec results in an orthorhombic model with a significantly altered oxygen distribution in the a-b plane as compared to the structure before relaxation. This supports the findings that the variation of the oxygen distribution in the a-b plane is a possible reason for the different crystal structures of $Ta_2O_5$. If we take a look at the $Ta_2O_5$ crystal structure proposed by Lehovec (Fig. 6), there are two different bonding types of the oxygen atoms. The oxygen atoms denoted with O1, O2, and O3 correspond to the same type, as they are always bonded to two tantalum atoms (Ta1-Ta1 or Ta2-Ta2 in c-direction; Ta1-Ta1 in b-direction), whereas O4 and O5 correspond to another type, as they are bonded to three tantalum atoms (Ta1-Ta2-Ta2). This configuration leads to different oxygen concentrations ($f_{ox}$, neighboring oxygen atoms divided by their coordination to tantalum atoms) of $f_{ox}$ = 2.66 around Ta1 and $f_{ox}$ = 2.33 around Ta2 atoms. In comparison, the a-b plane of the fully relaxed model shown in Fig. 7a and summarized in Table 2, shows the following oxygen distribution: O1, O2, and O5a/O5b are bonded to two tantalum atoms (Ta1-Ta1 or Ta2-Ta2 in c-direction; Ta1-Ta2 in a-b-direction), whereas O3 and O4 are three-coordinated to tantalum (Ta1-Ta1-Ta2 or Ta2-Ta2-Ta1). An important difference to the structure of Lehovec is the O5 atom, which flips from position O5a to O5b while going from one a-b plane to the neighboring one below (or above). As a consequence of this "flipping", see also Fig. 7b, this new arrangement of oxygen around tantalum results in the stoichiometric configuration of $f_{ox}$ = 2.5 for every tantalum and therefore, all tantalum positions are equal. For simplicity, this structure is called "FO-$Ta_2O_5$", which addresses the flipping nature of the O5 oxygen atoms.

As our deposited crystalline coatings are not stoichiometric, also sub-stoichiometric models with an O/Ta ratio of 2.33 were calculated. To obtain sub-stoichiometry either additional tantalum atoms or oxygen vacancies are needed. The formation of oxygen vacancies is energetically more favorable than the generation of interstitial tantalum atoms or the substitution of oxygen by a tantalum atom. Also the needed cell size for obtaining an O/Ta ratio of 2.33 is much smaller for the case of oxygen vacancies. For the case of additional interstitial tantalum atoms, 7 unit cells of the FO-$Ta_2O_5$ are necessary, whereas only 3 unit cells are necessary for the case of oxygen vacancies. Therefore, we have chosen three different oxygen deficiency models (3 times the FO-$Ta_2O_5$ in b-direction which corresponds to 6 times the formula unit). In the first case (I) two 3-coordinated oxygen atoms (O3 and O4), in the second case (II) two 2-coordinated oxygen atoms (O5a and O5b), and in the third case (III) one 2-coordinated (O5b) and one 3-coordinated (O3) oxygen atom were removed. O1 and O2 defects were not considered as they indicate stable positions for most of the published structures. *Ab*

*initio* calculations showed, that only the first case (I) having just 3-coordinated oxygen defects was stable. Consequently only this model was considered here.

Remarkably, calculated peak positions of the FO-$Ta_2O_5$ (nano-beam: 14.28, 14.45 deg., XRD $2\theta$ angles: 26.87, 27.57 deg.) and the FO-$Ta_2O_5$ with oxygen defects (nano-beam: 14.39, 14.37 deg., XRD $2\theta$ angles: 27.64, 27.59 deg.), would fit much better the 110 and 200 diffraction peaks at 0 deg. than the indexed peak positions of Lehovec (see Fig. 3). This is also valid for the 111 and 201 diffraction peaks at 45 deg. in Fig. 3, where the diffraction peaks of the *ab-initio* calculated FO-$Ta_2O_5$ (nano-beam: 18.72, 18.85 deg., XRD: 36.12, 36.38 deg.) and FO-$Ta_2O_5$ with oxygen defects (nano-beam: 18.86, 18.84 deg., XRD: 36.40, 36.36 deg.) would also fit much better. Together with the observed constant deviation from standard peak positions along the whole layer thickness, this indicates that the peak shifts are for a big part caused by the formation of FO-$Ta_2O_5$ and not based on e.g. biaxial stresses.

To further illustrate the differences between the various stoichiometric $Ta_2O_5$ structure models, we calculated their DOS. The band-gap of ~2.5 eV for our proposed FO-$Ta_2O_5$ model (Fig. 8a) is much closer to the experimentally obtained band-gap of around 4 eV [2,34,35] than the band-gap of ~1.5 eV obtained for the Stephenson model (Fig. 8b). At that point we note that the GGA calculated band-gaps are in general expected to be underestimated with respect to the experimental values, which is a well-known deficiency of standard DFT. The hexagonal (Fig. 8c) and the monoclinic (relaxed $U_3O_8$, Fig. 8d) structure yield even smaller band-gaps and the Lehovec model (Fig. 8e) hardly exhibits a band-gap. The latter is actually not stable during relaxation and thus we needed to fix the cell shape during relaxation. Introducing defects to our FO-$Ta_2O_5$ model results in no significant influence on the calculated band-gap width, see Fig. 8f, but results in electronic defect states in the middle of the band-gap, which was recently also observed for the Stephenson model [36]. Similar relaxations to our FO-$Ta_2O_5$ structure were recently published by Wu et al. [35], although they started with a structure published by Aleshina et al. [33], which leads to a deviation of $f_{ox}$ from 2.5 for every tantalum atom and a much smaller GGA band-gap of clearly below 0.5 eV.

The energies of formation ($E_f$) of the discussed structures (Table 1) exhibit with -3.259 eV/atom the smallest value for the supercell by Stephenson. Nevertheless, $E_f$ of our proposed FO-$Ta_2O_5$ structure with -3.209 eV/atom is by only 0.050 eV/atom higher. Furthermore, also our oxygen deficient FO-$Ta_2O_5$ model exhibits an energy of formation of -3.146 eV/atom, hence only 0.113 eV/atom above the Stephenson structure. The Lehovec structure at fixed cell shape ($E_f = -2.947$ eV/atom), the hexagonal ($E_f = -3.020$ eV/atom) as well as the monoclinic ($E_f = -2.962$ eV/atom) structures show considerably higher energy of formations and hence they are significantly less stable.

## 4. Summary and Conclusions

Structural investigations of reactively magnetron sputtered $Ta_2O_5$ films clearly show a crystallographic evolution from amorphous to crystalline growth with increasing film thickness. The crystalline parts are 110 – 200 textured and exhibit a highly strained lattice. Nano-beam diffraction across the whole layer thickness of ~15 µm show that certain crystallographic orientations exhibit a constant deviation from literature reports. As constant strains across a ~15 µm thin PVD film are very unlikely we have developed a new orthorhombic structure of $Ta_2O_5$ where one oxygen atom alters between two possible positions of neighboring unit cells. This so called flipping oxygen model (FO-$Ta_2O_5$) was proven by DFT calculations to fit the observed nano-beam diffraction and XRD peak-patterns. Furthermore, the energy of formation of -3.2098 eV/atom suggests that the newly developed structure is by far more stable than the Lehovec structure, the hexagonal as well as the monoclinic structures. The unique coordination of the oxygen atoms within our FO-$Ta_2O_5$ model guarantees that each Ta position is stoichiometric. Nevertheless, the accommodation of oxygen defects is energetically not very expensive and hence it can also describe slightly substoichiometric compounds. Moreover, the obtained band-gap of 2.5 eV for our FO-$Ta_2O_5$ model is by far closer to the experimental value of 4 eV than the band gap of other basic structure models.

As our model (with only 0.05 eV/atom higher in energy of formation) is almost as stable as the well accepted orthorhombic Stephenson superstructure, but easier constructed and as it provides a close fit to our diffraction patterns of the $Ta_2O_5$ films, we propose the FO-$Ta_2O_5$ model as a new alternative to describe $Ta_2O_5$.

**Acknowledgement**

This work has been supported by the European Community as an Integrating Activity 'Support of Public and Industrial Research Using Ion Beam Technology (SPIRIT)' under EC contract no. 227012. The financial support by the Austrian Federal Ministry of Economy, Family and Youth and the National Foundation for Research, Technology and Development is greatly acknowledged. The nano-focus end-station was equipped through financial support by the German Federal Ministry of Education and Research (BMBF, projects 05KS7FK3 and 05K10FK3) which is also greatly acknowledged. We thank the P03 team for provision of the beam to the nano-focus end-station.

## Tables:

Table 1: Lattice parameters of calculated and experimentally obtained structures of $Ta_2O_5$. For better comparability the lattice parameters are normalized to one formula unit.

| structure | crystal lattice | a [Å] | b [Å] | c [Å] | β [deg.] | V/atom [Å$^3$] | Energy of formation [eV/atom] | band gap [eV] |
|---|---|---|---|---|---|---|---|---|
| FO-$Ta_2O_5$ | ortho. | 6.425 | 3.769 | 3.853 | | 13.339 | −3.209 | 2.5 |
| Oxygen deficient FO-$Ta_2O_5$ | ortho. | 6.462 | 3.722 | 3.827 | | 13.811 | −3.146 | 2.5, (defect states) |
| Stephenson | ortho. | 6.270 | 3.759 | 3.824 | | 12.878 | −3.259 | 1.5 |
| Lehovec (fixed cell-shape) | ortho. | 6.297 | 3.724 | 3.947 | | 13.224 | −2.947 | 0 |
| Fukumoto [30] | ortho. | 6.357 | 3.670 | 3.890 | | 12.965 | −3.006 | 1 |
| | hex. | 7.341 | - | 3.890 | 120 | | | |
| monoclinic | mono. | 6.837 | 3.071 | 3.994 | 104 | 11.651 | −2.962 | 1 |
| Lehovec [27] diffraction file | ortho. | 6.200 | 3.660 | 3.890 | | 12.610 | | |

Table 2: Structural parameters of FO-$Ta_2O_5$ (a: 6.425 Å, b: 3.769 Å, c: 7.706 Å)

| site | x | y | z | occupancy |
|---|---|---|---|---|
| Ta1 | 0.00 | 0.00 | 0.00 | 1 |
| Ta2 | 0.45 | 0.41 | 0.00 | 1 |
| Ta3 | 0.00 | 0.91 | 0.50 | 1 |
| Ta4 | 0.45 | 0.50 | 0.50 | 1 |
| O1 | 0.98 | 0.95 | 0.25 | 1 |
| O2 | 0.46 | 0.45 | 0.25 | 1 |
| O3 | 0.11 | 0.47 | 0.00 | 1 |
| O4 | 0.34 | 0.94 | 0.00 | 1 |
| O5a | 0.72 | 0.20 | 0.00 | 1 |
| O6 | 0.98 | 0.95 | 0.75 | 1 |
| O7 | 0.47 | 0.45 | 0.75 | 1 |
| O8 | 0.11 | 0.44 | 0.50 | 1 |
| O9 | 0.34 | 0.97 | 0.50 | 1 |
| O5b | 0.72 | 0.70 | 0.50 | 1 |

Figure captions:

Fig. 1: Surface (a) and cross-sectional (b) SEM micrographs and XRD patterns (c) of reactively magnetron sputtered $Ta_2O_5$ thin films on silicon. Different deposition times of 5, 25, and 120 min, are indexed in these figures by 1, 2, and 3, respectively.

Fig. 2: Summed nano-diffraction pattern of all captured 2D diffraction pattern across the film thickness of the 120 min deposition.

Fig. 3: Evolution of the nano-diffraction patterns across the film thickness for the three cake-segments sketched in Fig. 2. The vertical lines are indicating the peak positions of the "Lehovec" structure.

Fig. 4: XRD pattern (logarithmic scale) of our coatings deposited for 120 min after annealing for 1 h up to 1000 °C in vacuum (a) and ambient air (b).

Fig. 5: Cross-sectional SEM micrographs of our coatings deposited for 120 min after annealing for 1 h at 1000 °C in vacuum (a) and ambient air (b).

Fig. 6: Schematic of the a-b plane for the $Ta_2O_5$ structure proposed by Lehovec et al. [7].

Fig. 7: Schematic of the a-b plane (a) and the c-b plane (b) for our *ab initio* calculated FO-$Ta_2O_5$ structure.

Fig. 8: *Ab initio* obtained DOS for our FO-$Ta_2O_5$ structure (a), the orthorhombic structure by Stephenson et al. [10] (b), the hexagonal structure by Fukumoto [30] (c), a monoclinic structure (relaxed $U_3O_8$) (d), the orthorhombic structure by Lehovec [7] (e) and our oxygen-deficient FO-$Ta_2O_5$ for obtaining an O/Ta ratio of 2.33 (f).

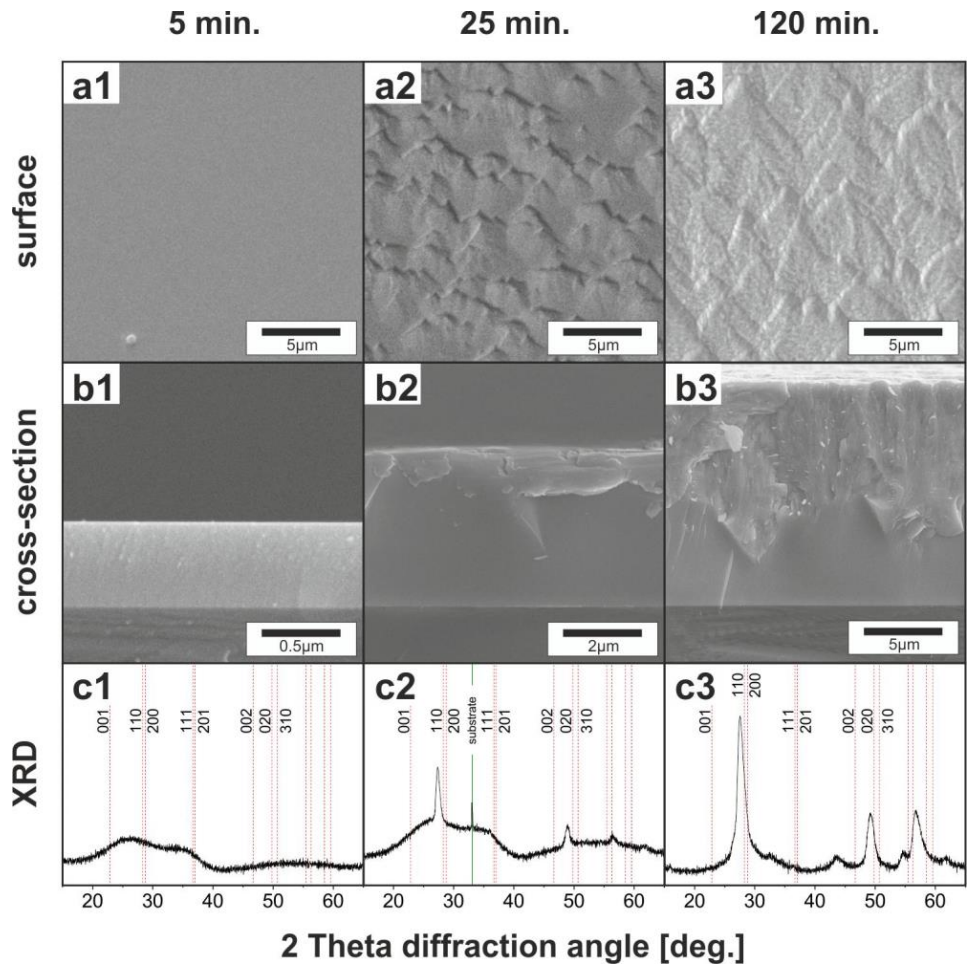

Fig. 1

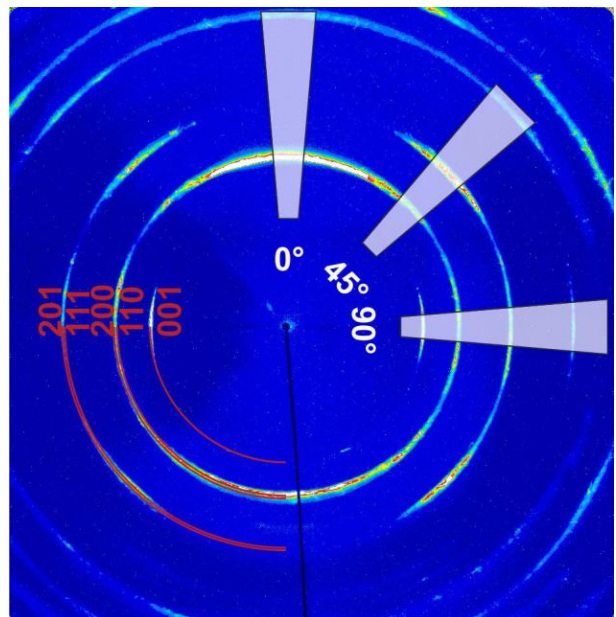

Fig. 2

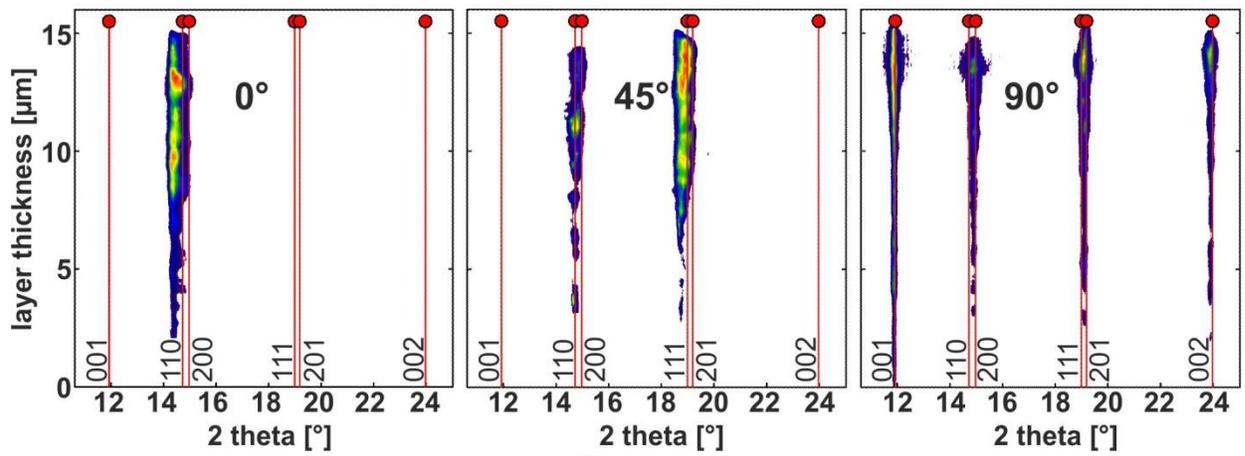

Fig. 3

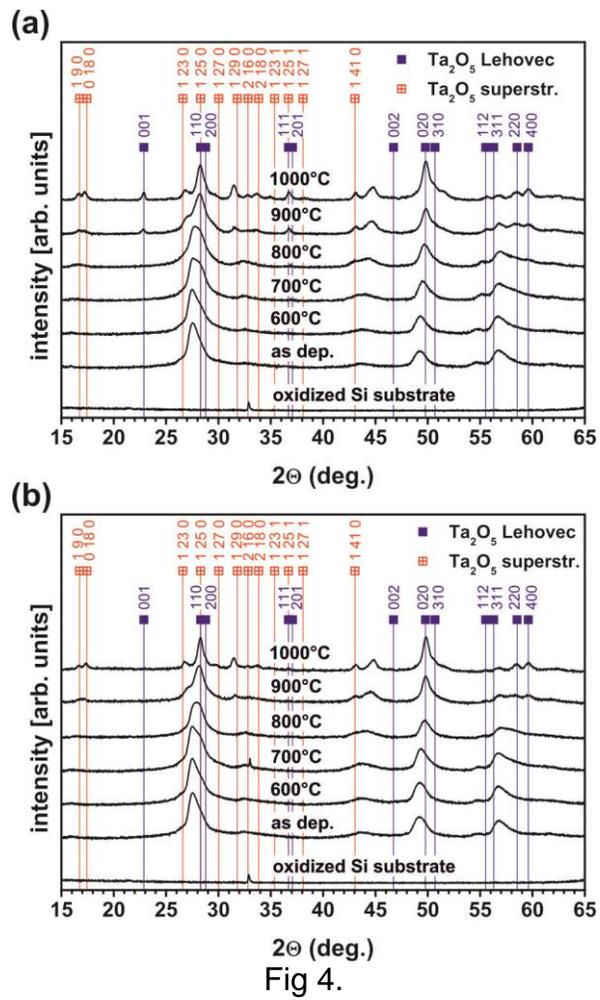

Fig 4.

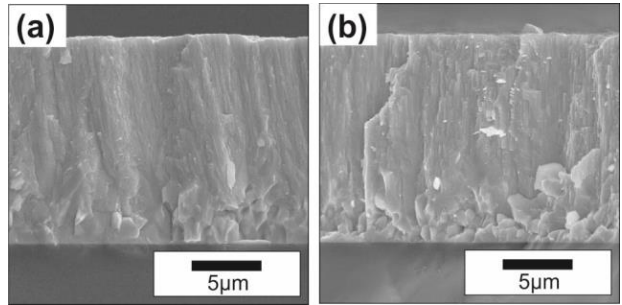

Fig. 5

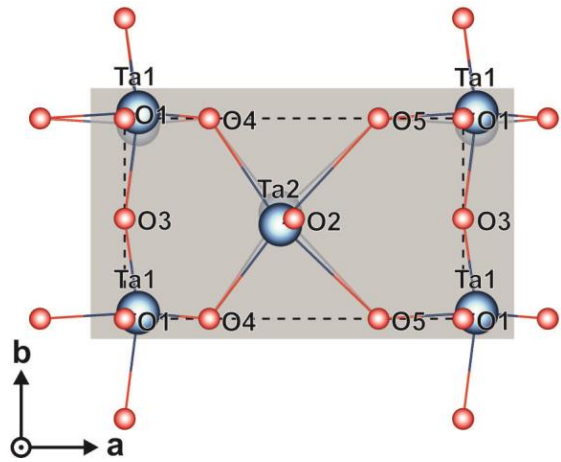

Fig. 6

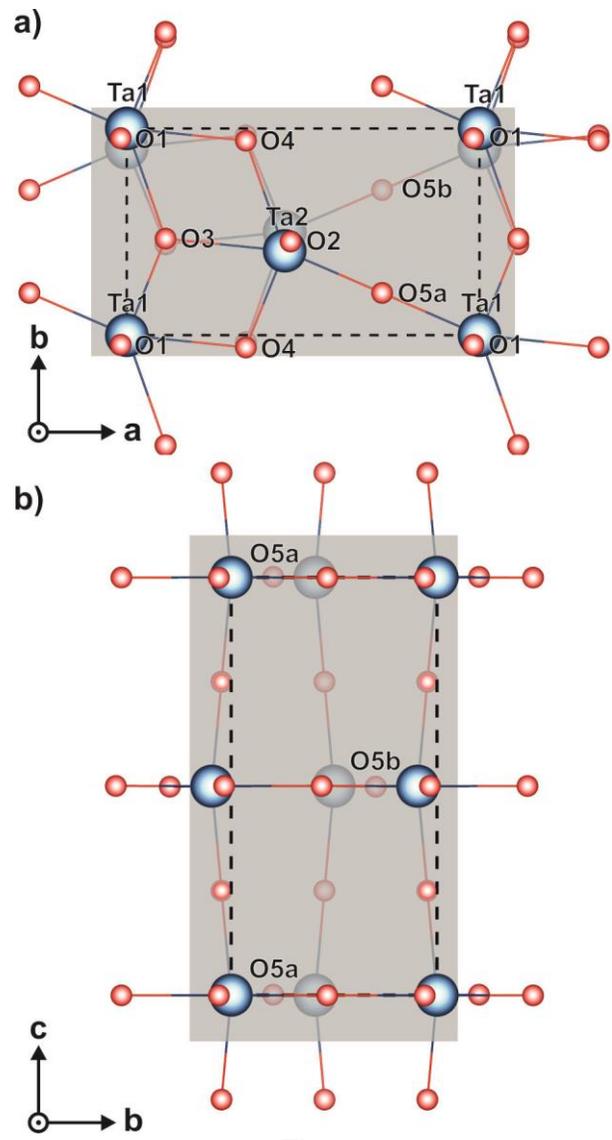

Fig.7

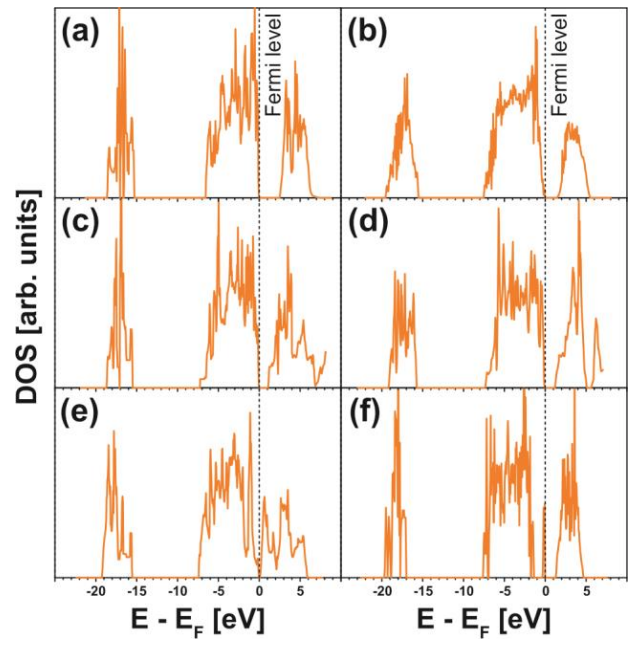

Fig. 8